\documentclass[aps,pre,twocolumn,showpacs,superscriptaddress,groupedaddress]{revtex4}
\usepackage{graphicx,subfigure}
\usepackage{hyperref}
\usepackage[svgnames]{xcolor}
\usepackage{amsmath}
\usepackage{amsfonts}
\usepackage{bm}
\usepackage{amssymb}
\usepackage{varioref}
\usepackage{float}
\usepackage{dcolumn}   
\usepackage{subfigure}
\usepackage{array}
\usepackage{tikz}
\usepackage{pgfplots}
\usepackage{mathtools}
\graphicspath{{./Figures/}}
\usetikzlibrary{positioning,fadings,through}
\definecolor{left} {HTML}{001528}

\begin{document}

\title{Structure and dynamics of a self-propelled semiflexible filament}
\author{Shalabh K. Anand }
\email{skanand@iiserb.ac.in}
\author{Sunil P. Singh}
\email{spsingh@iiserb.ac.in}
\affiliation{Department of Physics,\\ Indian Institute of Science Education and Research, \\Bhopal 462 066, Madhya Pradesh, India}

\begin{abstract}
We investigate structural and dynamical properties of a self-propelled filament using coarse-grained Brownian dynamics simulations. A self-propulsion force is applied along the bond vectors, i.e., tangent to the filament and their locations are considered in two different manners. 
In the case one, force is applied to all  beads of the filament, which is termed as homogeneous self-propulsion. 
Here, we obtain a monotonic decrease in the flexibility of the filament with  P\'eclet number. Hence, radius of gyration  also displays the same trend. Moreover, the radius of gyration of  the filament  
  shows universal  dependence for various bending rigidities with flexure number. The effective diffusivity  of  the filament shows enhancement with the active force and it increases linearly with force and bending rigidity.  
In the case two, self-propulsion force is applied only to few bond vectors. The location of active forces is chosen in a periodic manner starting from the tail  of the filament and leaving the front end without force.  
In this case, filament acquires various structures  such as rod-like, helical, circular, and  folded states. The transition from  several states is understood in terms of tangent-tangent correlation, bending energy and torsional order parameter.  The helical state is identified through a crossover from exponential to oscillatory behavior of the tangent-tangent correlation. A sudden increase in the bending energy  separates a helical to  a folded states of the filament. 
\end{abstract}
\maketitle

\section{Introduction}
Study of active matter systems such as a flock of birds \cite{vicsek1995novel}, school of fishes\cite{parrish1997animal}, bacterial colonies\cite{dombrowski2004self}, motility of spermatozoa etc., has drawn immense research interest in recent years\cite{Marchetti:2013,Elgeti:2015,Bechinger:2016}. Their movement is fuelled by  the chemical energy,  which is converted into mechanical energy. The presence of local excess energy drives the system out-of-equilibrium. Understanding  behavior of driven systems is an intense area of research from the fundamental aspect. Out of various driven or active macromolecular systems, one widely studied example is an active filament, which is  regarded as thin and long polymer chains\cite{Ghosh:2014, eisenstecken2016conformational,Eisenstecken:2017,Jiang:2014,Kaiser:2015,Raghu:2012,Holder:2015,Holder:2016,Laskar:2017,sarkar2016coarse}. Several types of active filaments are found in the cell, and they play a decisive role in providing shape, structure and motility to cell membranes\cite{howard2001mechanics}. Moreover, many 
microswimmers propel themselves by  long hairy polymeric  structures such as, cilia and flagellum\cite{brennen1977fluid,lyons2006reproductive,Elgeti:2015}. These active filaments exhibit interesting structural, dynamical and collective behavior\cite{rodriguez2003conserved,vicsek2012collective,abkenar2013collective, decamp2015orientational,sumino2012large,schaller2011polar,schaller2013topological}. 
 
In the recent past, various studies   have been done on the active filaments using theoretical\cite{Ghosh:2014,eisenstecken2016conformational,Eisenstecken:2017} and  simulation  models\cite{Ghosh:2014,Holder:2015,Holder:2016,Laskar:2017}. In these models,  either self-propulsion force is imposed  tangential to the filament\cite{chelakkot2014flagellar,Ghosh:2014,Holder:2015,Holder:2016,Laskar:2017} or monomers of the filament are treated as  active Brownian particles\cite{Kaiser:2015,eisenstecken2016conformational,Eisenstecken:2017}. A freely moving active filament acquires numerous  dynamical conformations, such as  rotational motion, snake-like motion\cite{Holder:2015,Holder:2016},   straight translational  motion\cite{Jiang:2014,sarkar2016coarse}, etc. 
 The rigidity of a filament plays a crucial role in its conformational behavior, as a flexible polymer swells under strong active force\cite{Kaiser:2015}, while a semiflexible filament shrinks under activity. However, in extreme propulsion limit  polymer  swells
again\cite{eisenstecken2016conformational,Eisenstecken:2017}. 
 A clamped filament shows  
  beating and spontaneous rotational motion under tangential 
  compressive force\cite{chelakkot2014flagellar,Laskar:2017}.   Presence of a load in front shows stable circular, beating and spiral structures on the surfaces in the absence of hydrodynamics\cite{Holder:2015,Holder:2016}. However, hydrodynamic interactions induce instability to a filament  when actuated with an active colloid on its terminus\cite{Laskar:2017}.

Internal relaxation of the filament is altered in the presence of active force. The change in relaxation time is determined by the strength of force and their correlations\cite{Nairhita:2016}. The longest relaxation time shows a crossover from bending dominated limit to the  flexible limit  under strong active force\cite{eisenstecken2016conformational,Eisenstecken:2017}. 
Active medium and the strength of  force influences the diffusive behavior of the filament. Interestingly, an active filament shows enhanced diffusion as well as the superdiffusive behavior. In a viscoelastic medium with active bath\cite{Vandebroek:2015,Osmanovic:2017},  mean-square-displacement 
of the centre-of-mass of filament as well as monomers display a super-diffusive behavior\cite{Ghosh:2014,Vandebroek:2015,Osmanovic:2017}. This dynamical aspect  of the filament has been reported in several experimental investigations in cellular medium, cytoskeletons and  chromatins \cite{maharana2012dynamic,talwar2013correlated,weber2012nonthermal,zidovska2013micron,brangwynne2008nonequilibrium}.

\begin{figure}
	\includegraphics[width=\columnwidth]{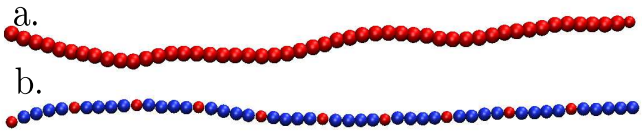}
 	\caption{Figure displays arrangement of the active monomers on the filament in red color and passive monomers are in blue color. Top filament (a) corresponds for the homogeneous self-propulsion and the bottom one (b) is for the periodic sequence of the active monomers. Distance between two successive monomers is $l_a=5$ (in simulation units).}
 	\label{Fig:pattern_active}
 \end{figure}

   In this article, we investigate a freely moving self-propelled filament in  bulk (three dimensions) using over-damped Langevin dynamics.  Self-propulsion force is imposed tangentially to   the filament. Two different cases are considered for the tangential self-propulsion. In the case one, active force is applied  homogeneously along the filament on each bond (figure~\ref{Fig:pattern_active}a). In the second case,  the role of inhomogeneity is considered in an averaged manner. Thus  the self-propulsion force is applied only to few
     bonds on the filament, and their location is chosen 
     in a periodic manner  as shown in  figure~\ref{Fig:pattern_active}b. This can be understood as a filament is divided into equal length of segments and active force is placed on the tail of each segment as shown in figure~\ref{Fig:pattern_active}b by red color. In this case, front monomers are passive which acts as load.   So far the role of periodic sequence of self-propulsion on a filament  has not been investigated in  detail.  Hydrodynamic interactions are ignored for the  simplicity
      of the calculation throughout the article.

   A freely swimming self-propelled filament buckles under the compressive force. We found that, the structural 
   change, under homogeneous self-propulsion, shows a universal behavior with respect to a dimensionless parameter called  flexure number for various bending  
   rigidities. In the periodic self-propulsion, a filament undergoes a transition from extended state to   a circular,  helical,  and folded or 
   strongly buckled states. Our main emphasis in this article is to identify 
    several phases emerging under various self-propulsion arrangement. To do so, we have calculated torsional order parameter 
   \cite{Raghu:2012}, bending energy  and tangent-tangent 
   correlation of the filament\cite{giomi2010statistical}.  
   In the extended state torsional parameter  is small, that becomes very large in helical and folded states. Furthermore, structural correlations along the filament exhibit  oscillatory behavior. Curvature radius of the helical phase shows a power law variation with force. 
       
  The article is  organised as follows. In  section-II, we discuss 
   a coarse-grained model for the self-propelled filament. Structural and dynamical behavior of the homogeneously 
   self-propelled filament is presented in section III. Results for the periodically arranged self-propelled monomers on the filament are discussed in section IV. Summary of the results is presented in section V.

\section{Simulation Model}
We model filament  as  a semiflexible linear polymer composed of a sequence of $N$ monomeric units connected via Harmonic spring. All the monomers in the polymer also interact via excluded volume interactions. The total potential  energy of the filament can be written as, 
$U = U_{h} + U_{b} + U_{LJ}$,
where $U_{h}$ is spring potential, $U_b$ is bending potential and  $U_{LJ}$  corresponds to excluded volume potential. Harmonic potential is given as, 
\begin{equation}
U_{h} = \frac{k_{s}}{2} \sum_{i=1}^{N-1}(|{\bm r}_{i+1} - {\bm r}_{i}|-l_{0})^2 ,
\end{equation}
 where $l_0$  is average equilibrium bond length, ${\bf r}_i$ is position vector of the $i^{th}$ monomer, and $k_s$ is the spring constant. Bending potential energy $U_b$,  which accounts for the stiffness of the polymer is written as,
\begin{equation}
U_{b} = \frac{\kappa}{2} \sum_{i=1}^{N-2}({\bm R}_{i+1}- {\bm R}_{i})^2,
\end{equation}
 here $R_i$ is length of the $i^{th}$ bond vector, defined as 
${\bf R}_i= {\bf r}_{i+1}-{\bf r}_{i}$, and $\kappa$ is the bending rigidity of the polymer which can be expressed in terms of   persistence length of the polymer $l_p$ as, 
$l_p=\kappa l_0^3/{k_BT}$, where $k_B T$ is thermal energy.

Excluded volume potential  avoids overlap  of beads  in  a polymer, and its form is taken from  truncated  repulsive part of Lennard-Jones potential, i.e., $R_{ij} < 2^{1/6}\sigma$,

\begin{equation}
U_{LJ} = \sum_{i=1}^{N-1} \sum_{j=i+1}^{N} 4 \epsilon
\left[\left(\frac{\sigma}{R_{ij}}\right)^{12}- \left( \frac{\sigma}{R_{ij}}\right)^{6} + \frac{1}{4} \right],
\end{equation}

and for  $R_{ij} \geq 2^{1/6}\sigma$, it is considered as, $U_{LJ}=0$. Here, 
$\epsilon$ is LJ interaction energy and  $\sigma$ is LJ diameter of the monomer. 

Newton's equation of motion for a monomer in overdamped limit is,
\begin{equation}
\gamma \frac{d {\bm r_i}}{d t} = - \nabla_{i} U + \textbf{F}_{r}^{i} + \textbf{F}_{a}^{i},
\label{eq:langevin}
\end{equation}
 where $\gamma$ is the friction coefficient, $\textbf{F}_{r}^{i}$ is the thermal noise  with zero mean, and $\textbf{F}_{a}^{i}$ is the self-propulsion force which is exerted on the $i^{th}$ bond vector. 
Since hydrodynamics interactions  are ignored in the simulations thereby solvent-mediated indirect 
coupling among the monomers is absent in the equation of motion.
 
 The viscous drag and the thermal noise  are related through the fluctuation-dissipation relation, 
 \begin{eqnarray}
\left< \textbf{F}_{r}^{i}(t).\textbf{F}_{r}^{j}(t')
\right> & = & 6k_{B}T\gamma \delta_{ij}\delta(t-t') 
\label{eqn-7}.
\end{eqnarray}

 Total self-propulsion force on the polymer is given as, 
\begin{equation}
\textbf{F}_{a}^T =\sum_{i=1}^{N}{\bf F}_a^{i}= \sum_{i=1}^{N-1} f_a {\bf t}({\bf r}_i) \Theta(r_i).
\end{equation} 

 Where $ {\bf t}({\bf r}_i)={({\bm r}_{i+1}-{\bm r}_{i})}/{ |{\bm r}_{i+1}-{\bm r}_{i}|} $ is unit tangent vector on  the $i^{th}$ monomer. Active force is shared equally between $i^{th}$ and $(i+1)^{th}$ monomers as $f_a/2$.  
If step function $\Theta(r_i) =1$, then $i^{th}$ monomer is active. Similarly  
 if $\Theta(r_i) =0$, then monomer will be passive. We consider here two different arrangements of self-propulsion force on the filament  described as follows.   
 i) Homogenous self-propulsion: Active force is applied on each bond of the polymer, therefore all monomers feel active force. We term it as homogenous self-propulsion.
  ii) Periodic sequence of self-propulsion: In this case, propulsion force is applied  to  few  bonds arranged in a  periodic sequence. 
  Placement of active monomers  starts from  the tail of  filament in equidistant manner with leaving  front end as passive. Fig.~\ref{Fig:pattern_active}b displays the clear picture of the periodic arrangement of active monomers on  the filament.

  In the periodic self-propulsion,  number of active monomers are taken as a variable parameter which varies from $N_a=0$ to $N_a=N$. Here $N_a=0$ corresponds to the passive filament, while $N_a=N$ recovers the homogeneous 
  self-propulsion. Strength of the active force is defined in units of thermal energy called as P\'eclet number, which is a dimensionless number. It is defined here as,    
  $Pe = \frac{f_{a} l_0}{k_{B} T}$. 
  In the limit of $Pe<<1$, thermal fluctuation 
dominates. However, in the limit of $Pe>>1$, active force  dominates. 
 The ratio of the P\'eclet number ($Pe$) and the scaled persistence length ($l_p/l_0$) is  a dimensionless parameter given as, $ \chi=l_0Pe/l_p$. Here, $\chi$ is called flexure number, which provides a measure of active force over the bending rigidity.
 Flexure number is used to understand the buckling instabilities\cite{sekimoto_buckling,chelakkot2014flagellar}, spontaneous spiral formation, spiral stability  and rotational motion\cite{Holder:2015,Holder:2016} of active filaments.  

All the physical parameters presented  here are scaled in units of the bond length $l_0$, diffusion coefficient of a monomer $D_m$,  and thermal energy $k_{B}T$.  
 Simulations parameters are chosen as, 
  $k_{s}=1000k_{B}T/l_{0}^{2}$, $\sigma=l_0$,
   $\epsilon/k_{B}T=1$, and  time is in units of $\tau = l_0^2/D_m$, and stiffness parameter $\kappa$ is in units of $k_{B}T/l_{0}^{2}$.  Separation between active sites are considered in the range of $l_a=1$ ($N_a=101$) to $l_a=25$ ($N_a=4$) and the
    bending rigidity of the polymer is varied in the range of 
    $\kappa=0$ to $200$ for  homogeneous self-propulsion, whereas for patterned case we take $\kappa=100$ and $\kappa=40$. Unless explicitly mentioned, number of monomers in the chain is taken to be $N=101$. 
    We use Euler integration technique to solve Eq.~\ref{eq:langevin}. The integration time step 
  $\Delta t$ is varied from the range of $10^{-2}\tau$ to $10^{-5}\tau$ to ensure stable simulation results.
  All the simulations are performed  in cubic periodic box in three spatial dimensions.

   All simulations are restricted in the range of P{\'e}clet number $0$ to $600$. A larger active force causes increase  in the average bond length  thus it requires larger spring constant. In addition to that, smaller integration time step needs to be taken into account for better numerical accuracy.  
  In order to ensure better statistics, each data point  is averaged over  $50$  independent runs in small P{\'e}clet number limit,i.e., $Pe<1$. However, rest of the data points are averaged over 32 independent runs.

\section{Homogeneous Self-Propulsion}
In this section, we present results for homogeneous self-propulsion where all monomers are 
active. In equilibrium, structural and dynamical behavior of the filament  is very well understood in the past\cite{winkler1994models,harnau1997influence,
KRATKY194935,Saito:1967,Hsiao:2010,Hsia0:2011}. 
The presence of propulsion on the filament causes bending, therefore its structural and dynamical behavior deviates from the equilibrium.    

\subsection{Structural Properties}
\begin{figure}
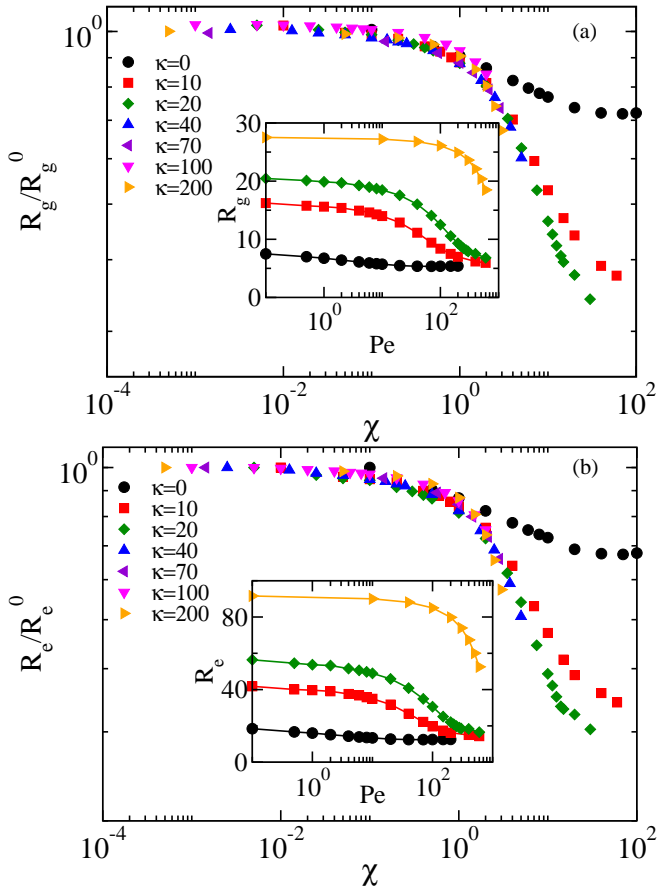

\includegraphics*[width=\linewidth]{struc_homogen_rg.eps}\\\includegraphics*[width=\linewidth]{struc_homogen.eps}
	\caption{Figure shows the relative change in the mean radius of gyration  $R_g/R_g^0$ (Plot a) and  the end-to-end distance  $R_e/R_e^0$ (Plot b) w.r.t. its equilibrium values $R_g^0$ and $R_e^0$ respectively  as a function of flexure number $\chi$. Inset of figure a and b shows the mean radius of gyration and the mean end-to-end distance $R_e$ as a function of $Pe$.  Various curves in figure a and b corresponds  to $\kappa$ as displayed in the figures.} 
	\label{fig:rg_re}
\end{figure}

The structural change is analysed by estimating average radius of gyration 
$R_g$, and  average end-to-end distance $R_e$ of the filament. The radius 
of gyration and the end-to-end distance is computed by the expression, 
\begin{equation}
R_g^2 = \frac{1}{N}\left<\sum_{i=1}^{N} ({\bm r}_i-{\bm r}_{cm})^2\right> ~;~ R_e^2= \left<({\bm r_1}-{\bm r_N})^2\right>,
\end{equation}
where ${\bm r}_{cm}$ is the centre-of-mass of the filament and angular brackets indicate the ensemble average of the physical quantities.  
 Figure~\ref{fig:rg_re} shows the dependence of  radius of gyration and end-to-end distance  on flexure number  for various bending rigidities.
  
 First, we discuss results of the flexible polymer, i.e., for $\kappa=0$.   In the weak active force,i.e., $\chi<<1$, the normalised radius of gyration $R_g/R_g^0$ decreases  monotonically  with increasing $\chi$. Further increase in  $\chi$ results relatively large change in $R_g/R_g^0$ and eventually in  the  limit of large $\chi >10$, $R_g$ is almost independent of $\chi$, which is consistent with other simulations and theoretical  results\cite{Eisenstecken:2017,eisenstecken2016conformational}. The saturation in $R_g$ and $R_e$  occurs  due to excluded volume interactions  which  prevents collapse of a polymer.

Now we present results for the semiflexible filament.  
Here, the radius of gyration decreases with activity as 
displayed in figure \ref{fig:rg_re}. In the limit 
of weak active force, i.e., $\chi<1$, the radius of gyration gradually decreases with $\chi$ similar to the flexible polymer. Further increase in active force, i.e., $\chi \geq 1 $, the relative change in the radius of gyration increases substantially. 
Interestingly, for $\kappa>10$, all the  curves display nearly a universal behavior  up to four orders of  magnitude in $\chi$  as shown in  figure \ref{fig:rg_re}. The end-to-end distance also reflects the same  universal trend whose properties can be  understood in terms of a single master curve in limit of 
$\chi < 10$. 
 In the semiflexible regime, $R_g$ and $R_e$ decreases monotonically with $\chi$ up to $\chi \sim 1$. For $\chi >>1$, $R_g$ approaches to a plateau value for $\kappa=10$ and $20$ similar to the flexible polymer limit as reflected in  Fig.~\ref{fig:rg_re}a and \ref{fig:rg_re}b.  
 The inset of Fig.~\ref{fig:rg_re} a and b display  absolute values of $R_g$  and $R_e$ as a function of $Pe$ for various $\kappa$. This reflects the monotonic decrease of $R_g$ and $R_e$  in the range of $Pe \le 10^2$.  For the $\kappa \le 20$,  $R_g$ and $R_e$ are slowly approaching to a  saturation limit. Note that for $\kappa > 20$,  the saturation limit is not reached in displayed simulation range, which will appear in the limit of  $Pe > 10^3$.

A filament shrinks under  homogenous self-propulsion due to several competitive forces on it, which   can be understood  in terms of fluctuations in bond orientations due to random motion of monomers. As  a bead always goes to random motion which causes fluctuations in  bond orientations along the contour. Hence tangential force pushes filament along the randomly fluctuating bond  directions, which leads to increase in the average noise over the filament. Thus it bends and buckles with $Pe$,  which results in the  shrinkage of the filament. Considerably larger values of $\kappa$ suppress bond fluctuations, thus shrinkage of the polymer is smaller, even for the substantially large  P\'eclet numbers. 
  
\begin{figure}
\includegraphics*[width=\linewidth]{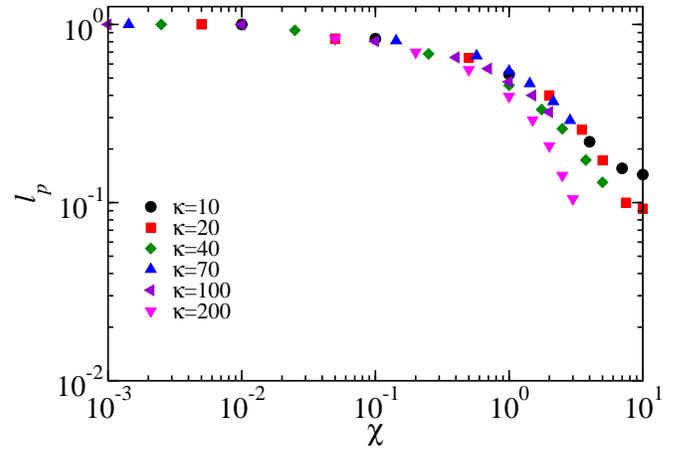}
	\caption{The persistence length $l_p$  as a function of $\chi$ for various $\kappa=10,20,40,70,100$ and $200$ for the polymer length $N=101$.}
	\label{fig:pers}
\end{figure}

  The shrinkage in the radius of gyration with activity reflects a decrease in  the rigidity of the polymer. To quantify the change in the rigidity,  we compute its persistence length with active force. This can be computed from the  tangent-tangent auto-correlation of the filament, which is expressed as, 
  $C(s)=< {\bf t}({\bm r_i})\cdot  {\bf t}({\bm r_{1})}>$,  where  $s$ is the arc length, $s=|r_i-r_1|\equiv |i-1|$. We have estimated correlation from one end of the polymer, i.e., from the tail for  $i=1,2...N-1$. The correlation decays exponentially as, $ C(s)\sim \exp(-\frac{s}{l_p})$,  with the arc length of the polymer  for all values of  $Pe$ and $\kappa$.  We calculate  persistence length  by fitting exponential function in the correlation function. Figure~\ref{fig:pers} displays the persistence length as a function of $\chi$, as expected,  $l_p$ decreases monotonically  for all $\kappa$ in the limit of   $\chi \leq 1$.  
  In the limit of
  $\chi>1$,  active polymer becomes flexible thereby persistence length, $l_p \approx l_0$, thus in the activity dominated regime $\chi>1$, $l_p$  nearly saturates  as displayed in figure~\ref{fig:pers} for smaller $\kappa=10$ and $20$. 
  In the limit of  $\chi \leq 1$, all the curves display a universal behavior with $\kappa$ as illustrated  in the plot. The analysis of the persistence length suggests that a semiflexible polymer  behave as a flexible polymer in the limit of large $\chi > 10$.

\subsection{Diffusion of Filament}
\begin{figure}
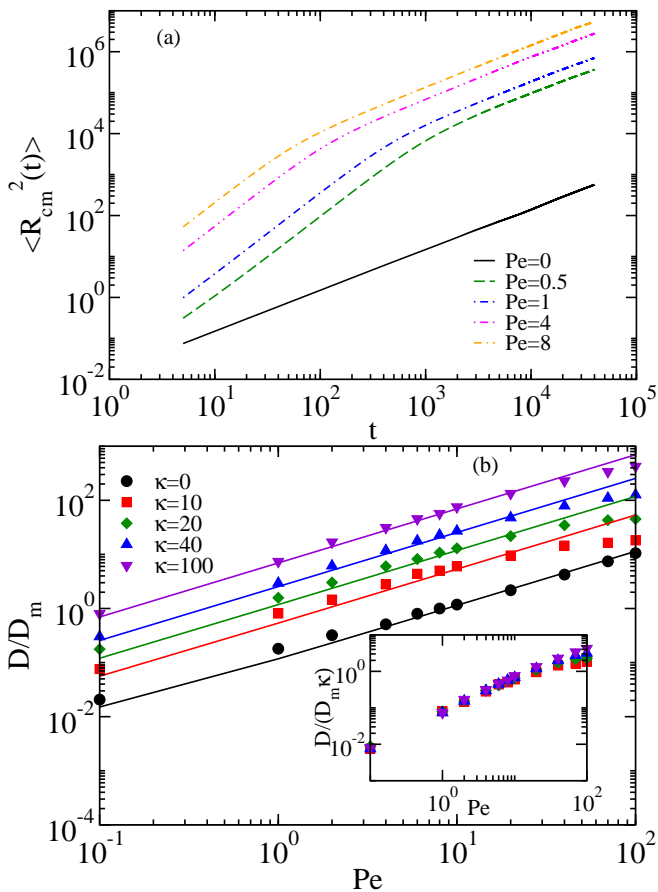

	\includegraphics*[width=\columnwidth]{news6}\\
	\includegraphics*[width=\columnwidth]{diff}	
	\caption{a) The mean-square displacement of the centre-of-mass of the filament for $\kappa=40$ for various $Pe=0.0,0.5,1,4$ and $8$. b) The scaled effective diffusivity $D/D_m$ of the filament as a function $Pe$ for various $\kappa=0,10,20,40$ and $100$. Inset shows scaled diffusivity $D/(D_m\kappa)$ with $Pe$ for $\kappa=10,20,40$ and $100$.}
	\label{fig:msd_hom}
	\end{figure}

In this section, we discuss  dynamics of a filament under homogeneous self-propulsion by estimating mean-square-displacement(MSD) of the centre-of-mass.  The MSD of a filament is computed as, $<R_{cm}^{2}(t)>=<[r_{cm}(t)-r_{cm}(0)]^{2}>$, where angular bracket denotes ensemble average.  In equilibrium, qualitative behavior of the MSD of a filament can be separated in mainly two time regimes.  
The short-time  limit is called ballistic regime, where $<R_{cm}^{2}(t) >\sim t^2$, however  long-time limit is  called diffusive regime, here the MSD varies linearly in time given as, $<R_{cm}^2(t)> = 6 Dt $, where $D$ is  the diffusion coefficient of the filament. Fig.\ref{fig:msd_hom}, displays the MSD of the filament with  time for different $Pe$. The MSD of the passive filament is shown from the solid line. As expected, it reflects diffusive behavior in the long-time limit.

 In the presence of tangential active force, the polymer is compressed and drifted along the end-to-end vector. Drift motion reflects  the enhanced mean-square-displacement of the centre-of-mass in the intermediate time regime. Increase in the MSD  is observed with  $Pe$ in all  time regimes. It is evident from the plot, the MSD exhibits  superdiffusive  behavior,i.e., $<R_{cm}^2>\sim t^{\alpha}$, with exponent  $\alpha > 1$,  in the intermediate time regime. In the long time limit, the MSD  recovers  the linear  behavior. Another interesting observation is the shift in  timescale of   
 super-diffusive regime  to smaller time  with P\'eclet number. This can be interpreted in terms of  the flexibility and memory of directed motion.  A filament  becomes flexible with active force as shown in the structural analysis, which results decrease in the persistence length of the directed motion. Thereby, superdiffusive behavior  persists relatively at smaller time. In a more intuitive way, we can also derive an approximate expression for the crossover time by equating the MSD expression in ballistic  regime $<R_{cm}^2(t_{c})> \sim (f_a t_c/\gamma)^2$, and diffusive regime $<R_{cm}^2(t_{c})> \sim  D t_c$, which gives $t_c \sim D \gamma^2/f_a^2$.

 The long time diffusivity of  the filament is obtained  by fitting a linear function to the MSD curves.
 Figure~\ref{fig:msd_hom}b displays diffusivity  of the polymer  with $Pe$ 
 for  several values of the rigidities. The diffusivity enhances   with  the P\'eclet  number. Moreover, it  increases linearly  with $Pe$, as, $D\sim g(\kappa)Pe$, where $g(\kappa)$ shows  bending dependence of $D$, for all $\kappa$ as illustrated in Fig.~\ref{fig:msd_hom}. The effective diffusivity of the filament increases  due to active fluctuations in the system. In an active medium,  a  modified form of the fluctuation-dissipation-relation(FDR) is proposed that relates the measured effective diffusivity of  a tracer particle with effective temperature as, $D=k_{B} T_{eff}/\gamma$~\cite{loi2011non,PalacciPRL2010}. We use above relation to define the effective temperature $T_{eff}$ in terms of long time diffusivity of the filament.   
  Thus  the effective temperature of the filament increases with active force as also reported in Ref.~\cite{Ghosh:2014}.
 
  The crossover time $t_c$ defined above can be re-expressed in terms of  $Pe$ and effective diffusivity as $t_c \sim D/Pe^2 $, linear dependence of $D$ on $Pe$ gives $t_c \sim g(\kappa)/Pe$. Inset of Fig.~\ref{fig:msd_hom}b  shows that $D/(\kappa D_m) \sim Pe $,i.e., $g(\kappa)\sim \kappa$. Hence, we obtain,  $t_c \sim  \chi^{-1}$. Thus the crossover time decreases  with increasing $\chi$,  our simulation also shows  decreases in $t_c$  with $\chi^{-1}$. 
 
The linear  form of the diffusivity of a filament with $Pe$ near the surfaces is also reported in previous studies\cite{Holder:2015,Holder:2016}. 
 At large bending parameters, persistence length of the directional motion  increases therefore the diffusivity also increases, which shows linear increase as expected with $\kappa$ for a given P\'eclet number.  Thus, a stiffer filament diffuses relatively faster than the flexible polymer under homogeneous self-propulsion.

\section{Periodic Sequence of Active Force}
 In this section, we explore various conformations 
 of the filament under equally spaced active monomers. Leaving front monomers as passive, which  acts as a load, leads to 
  bending  of the filament under compression. Presence of activity on the front monomers pull the front beads, which suppress the buckling between front active monomers. In addition to this, it also drags the filament 
 along the same direction that leads to translation motion. Thereby, front monomers are always left as passive. Importance of load is discussed and analysed in the conclusion section for the case of inhomogeneous self-propulsion.
  
   The arrangement of the active force on the filament is shown in Fig.~\ref{Fig:pattern_active} for the  separation $l_a=5$. Here, $l_a$ is the distance between two successive active monomers.   Interestingly, under a periodic sequence of active force,  a filament acquires interesting conformations during motion which is not observed in homogenous self-propulsion in bulk.  These conformations depend on the spacing between active forces $l_a$, the strength of propulsion and the  rigidity of the filament.
  
 In small P\'eclet number $Pe<1$, filament  translates along the direction of end-to-end vector, thus  its structure is weakly perturbed. In the intermediate regime of  compressive force, external force becomes comparable to the elastic energy. Therefore an active monomer pushes against passive monomers, which causes bending of the filament. Further increase in the force  buckles the filament in a  correlated manner throughout the filament.  This occurs for the separation of the active monomers $l_a\geq 5$.  In case of  correlated buckling, it forms a circular or  helical  structure. 
    Circular phase appears in the range of $l_a\geq 14$  and narrow range of $Pe$. In our analysis, a circular  phase is treated as  helical phase. This can be  visualised as a circular structure  is  a helix  of one helical turn with very small pitch length. 
 Further increase in the force pushes filament strongly against the viscous drag, which causes uncorrelated buckling and it leads  to distortion in the structure. Large force results to  sharp bending  and twisting of the filament nearby the active monomers, such sharply bent structures are called folded structures here.

 Few snapshots of the filament in the extended, helical, circular and folded states are shown in figure~\ref{Fig:snapshot}. We have also shown these structures in supplementary movie files in $S1$, $S2$ and $S3$.
   Our focus in this section is to identify helical, circular and folded states of a self-propelled filament in the parameter space of $Pe$ and $l_a$. To do so, we compute two-point tangent-tangent correlation of the bond vectors, the bending energy and the torsional order parameter of the filament.  
 \begin{figure}
 	\includegraphics[width=\linewidth]{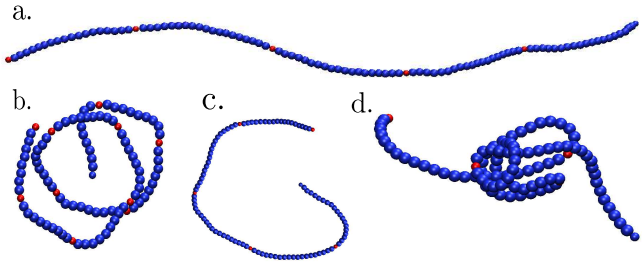}
 	\caption{Few snapshots of the filament showing different structures under periodic arrangement of active monomers. Red color corresponds for the active monomer and blue color represents for the passive one. Top row represents for the extended state (a). In the bottom row helical(b), circular(c) and folded structures(d) are shown form left to right respectively. }
 	\label{Fig:snapshot}
 \end{figure}

\subsection{Helical Phase}
\begin{figure}
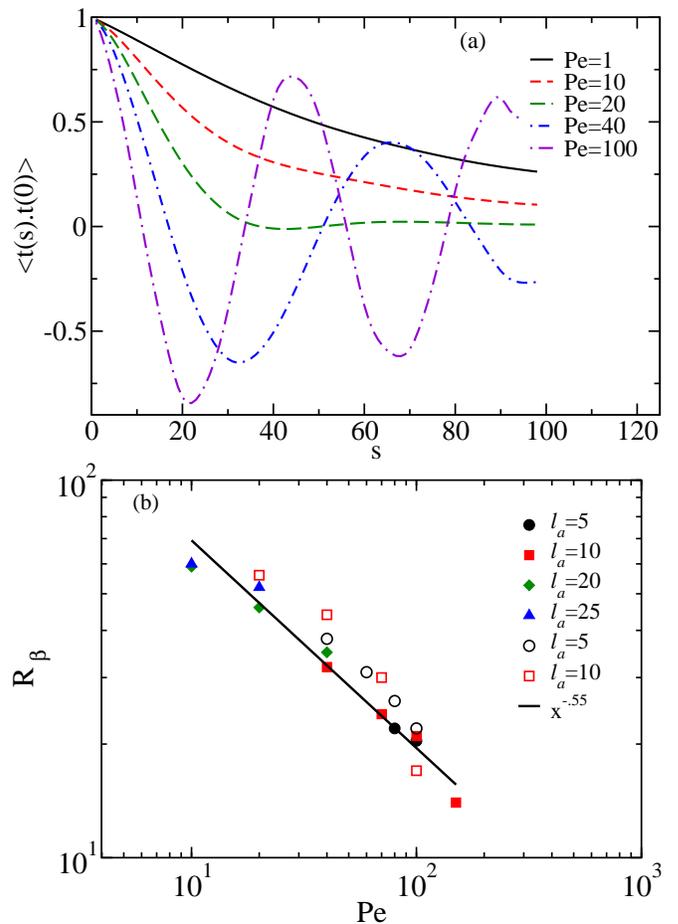

	\includegraphics*[width=\columnwidth]{tan_corr.eps}
	\includegraphics*[width=\columnwidth]{radius_fin.eps}
	\caption{a) Figure displays tangent-tangent  
	 correlation function $<{\bf t}(s)\cdot {\bf t}(0) >$ of the self-propelled filament at $l_a=10$ and bending rigidity $\kappa =100$.
	 b)  Figure displays curvature radius $R_{\beta}$ of the helical phase  obtained from fitting  Eq.~\ref{tangent} for various $l_a=5,10,20$ and $25$ for $\kappa=100$ (filled symbols). Open symbols corresponds  for $l_a=5$ and $10$ at 
	 $\kappa=40$. The solid line shows  power behavior $\left({Pe}\right)^{-.55}$ of the curvature radius.}  
\label{coiled}
\end{figure}

A helical phase can be distinguished using tangent-tangent correlation of the filament. The characteristic behavior of helical state is function of $l_a$ and $Pe$. 
 To quantify the helical state in above parameter space, we compute  tangent-tangent correlation function of the filament as discussed earlier.  Fig.~\ref{coiled} shows variation of the correlation for  few P\'eclet numbers at $l_a=10$. Here the correlation is computed from the end monomers as a function of arc length $s$.  The correlation function $C(s)$ decays exponentially  in the weak force limit with $s$. For larger force,i.e.,$Pe=20$, correlation  sharply reaches to  negative values after approaching to a minimum  it tends to zero from the negative side.  Further increase in the $Pe$ shows oscillations in the correlation. At larger $Pe$, these oscillations show
 long-range correlation. Strong correlation and sinusoidal  behavior signify the underlying helical phase of the filament\cite{giomi2010statistical}. 
Note that, under homogenous self-propulsion oscillatory behavior in the correlation is absent throughout the parameter regime presented in this article. Hence, we conclude that  no helical structure occurs in this case. Therefore, crossover from the exponential to sinusoidal correlation is the signature of  a helical phase. Sinusoidal behavior of correlation  occurs in  both  circular and helical phases.

A characteristic length scale and  wave number associated with the correlation can be estimated from the following expression,
\begin{equation}
C(s) = a_{\beta} \exp(-s/l_{p})(\cos( 2\pi s/R_{\beta}),	
\label{tangent}
\end{equation}
 where $a_\beta$ is some constant and $R_\beta$ is the length scale associated with the curvature radius of the filament. This function exhibits the property that it decays  exponentially and captures oscillatory behavior also.
 Fitting  Eq.~\ref{tangent} in the tangent-tangent 
correlation gives the curvature of the filament, which is  plotted in Fig.~\ref{coiled} as a function of $Pe$.  As expected, curvature radius decreases with increasing compressive force. Interestingly, $R_\beta$ for all the $l_a$'s shows a similar trend   and it follows a master curve without any scaling parameter as shown in the graph. Moreover, the master curve exhibits a power law behavior with  P\'eclet number, as $R_\beta\sim Pe^{-\beta}$, with the exponent $\beta \sim 0.55$.  The solid line in  Fig.~\ref{coiled} is shown for visualisation of the power law behavior with the same exponent.  The curvature radius of the filament obtained here has slightly larger exponent compare to a buckled filament near 
the surfaces\cite{Raghu:2012}, which  we believe due to larger contour fluctuations in the bulk relative to the surfaces.

\begin{figure}
\includegraphics*[width=\linewidth]{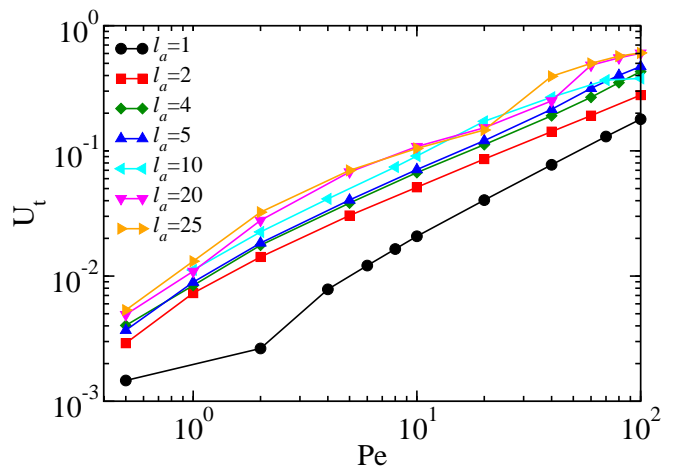}\\
\caption{Average torsional order parameter $U_t$ as a function of $Pe$ for various $l_a$ as shown in the figure for a given $\kappa=100$.}
\label{torsion}
\end{figure}

To show  out of plane motion  and twist of the filament in the helical phase, we compute average torsional order parameter of the filament. This is computed  in a manner similat to that discussed in Ref.~\cite{Raghu:2012}, 
\begin{eqnarray}
U_{t} &=& \sum_{i=2}^{N-1} \frac{1}{N-3}\cos{\theta_{i}} , \\ \nonumber
\cos{\theta_{i}} &=& \frac{(\vec{R}_{i-1}\times \vec{R}_{i}).(\vec{R}_{i}\times \vec{R}_{i+1})}{|\vec{R}_{i-1}\times \vec{R}_{i}||\vec{R}_{i}\times \vec{R}_{i+1}|},
\end{eqnarray}
 here $U_t$ is average of the sum of cosines of the torsion angle over the filament
and  $\theta_{i}$ is  the angle given by three consecutive bond vectors 
$\vec{R}_{i-1}, \vec{R}_{i}$ and $\vec{R}_{i+1}$. 
Figure~\ref{torsion} displays $U_{t}$  with  P\'eclet number for various $l_a$. The torsional order parameter smoothly increases with  $Pe$.  Figure~\ref{torsion} displays $U_t$ that is very small in the limit of $Pe << 1$, further increase in $Pe$ leads  to a monotonic increase in the torsional order parameter. In the limit of large $Pe$, $U_t$ has increased nearly two orders of magnitude. Large values of $U_t$ signifies that filament can be strongly twisted or folded, but  it's  not sufficient enough to delineate the helical structure. The combination of torsional order parameter and tangent-tangent correlation can precisely define the helical state.  
It would be important to mention here  that torsional angle does not reflect any sharp transition from  extended to helical and later state to folded state. Rather it varies smoothly from one phase to another phase with  force\cite{helical_kemp_1998}.

Another interesting observation worth mentioning here is the increase in the torsional parameter with $l_a$ for a given $Pe$. Fewer active monomers produce  large amount of twist or out of plane motion to the filament. In the limit of small separations, translation motion is dominant thus the change in torsional parameter is small. Increasing separations cause a larger drag force on the active monomers and thus reduce the translation motion, which buckles  the filament between active locations. This is reflected in the increase in torsion order parameter, from this we infer that in order to have a correlated buckling large separation among the active monomers is essential.

\begin{figure}
\includegraphics*[width=\linewidth]{bending_energy.eps}\\
\includegraphics*[width=1.1\linewidth]{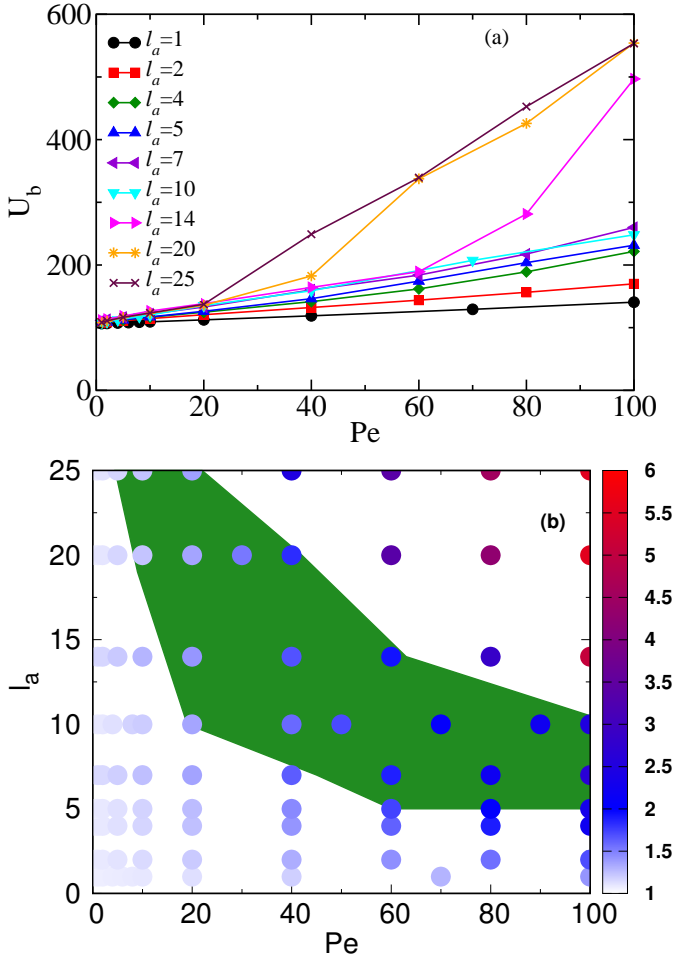} 
\caption{
a) Average bending energy of the filament as a function of $Pe$ for various spacing between active monomers. 
b) Phase diagram for the equispaced activity on the polymer for a given $\kappa=100$. The light blue circle  on the graph shows extended structure, green shaded area represents the helical and circular structures. Above the green shaded area the folded state is represented. In order to accomodate equal spacing between the active monomers  we have taken filament length $N=99$ for the $l_a=7$ and $14$.}
\label{phase_dia}
\end{figure}

\subsection{Folded state} 
In this section, we quantify uncorrelated buckling of the filament. 
In the limit of large force,  helical or circular states are followed by folded state. In this structure, a filament is strongly folded near the active monomers due to large buckling. To distinguish the helical state from the folded state, we compute the bending energy of the filament.  Figure~\ref{phase_dia}a displays variation of  bending energy of  the filament with  $Pe$. Bending energy increases linearly with  $Pe$ in the range of $l_a<14$. For the $l_a>14$, bending energy increases monotonically  up to a certain P\'eclet number, after which an abrupt change in the bending energy appears specifically for $l_a=14,20$ and $25$ as shown the Fig~.\ref{phase_dia}a. The transition point from the helical to the folded state is recognised from the sharp increase in the bending energy. 

  A sudden increase in the bending energy arises from the sharp buckling of the filament, which is  uncorrelated over the length scale of $l_a$. Therefore, sharp buckled filament exhibits very high bending energy.  Once again it's important to mention here that elastic or bending  energy does not show any sharp transition between the  extended state to the helical or circular state. However, a folded state does show sharp increase in the elastic energy. Furthermore, bending energies of the circular  and the helical states exhibit similar behavior. 

From the analysis of  torsional order parameter, the bending energy, and the tangent-tangent correlation  different structures of the filament in the parameter space
 of $Pe$ and spacing between active forces  are recognised. A phase diagram is  displayed in the Fig.~\ref{phase_dia}b as a function of $Pe$ and $l_a$. A color map shows the variation in bending energy per monomer in different phases, similarly color of the symbols are also changed in the plot. In graph, extended state and uncorrelated folded state are separated  by the green shaded  area which reflects the helical and circular states.  
 
 In the green shaded area, a circular phase appears  for $l_a =14,20 $ and $25$ and a helical is for $l_a=5,7$ and $10$.  We are analysing various structures in terms of bond correlation, torsional energy, and bending energy. These physical quantities do not show any significant change  from a circular to helical states as displayed in figures \ref{coiled}, \ref{torsion}, and \ref{phase_dia}a. Thus in the phase diagram, a circular state is displayed together with the helical state. The region below the green shaded area  represents the extended state, and above the shaded area corresponds to the folded state. Below the green area an extended phase appears, which occurs in smaller separation limit $l_a \le 4$ for all $Pe$.  In addition to that, for large separations  extended state appears  in the limit of small P{\'e}clet number. 
  
\section{Summary and conclusions}
 In this article, we have performed a detailed  study of the structural and dynamical behavior of a freely moving 
  self-propelled  filament in three dimensions. The location and the number of self-propelled monomers are considered as a variable, and they are mimicked in two different ways, homogeneously, i.e., on all monomers and  periodic sequence of  the self-propulsion on the filament. 
 
 In the homogeneous self-propulsion,  the radius of gyration and the end-to-end distance  decreases with increasing active force for all the bending parameters.  We have shown that  the change in  structural properties for various rigidities can be described by flexure number with a single master curve in the limit of $\chi \le 1$.  Similarly, our analysis also reveals that the persistence length of the polymer decreases with active force, a universal behavior for all  bending parameters is also  shown in the limit of $\chi  \leq 1$. 
The translation motion is dominant for $\chi<<1$, thus it acquires the extended state. In the limit of $\chi \approx 1$ the  polymer gradually enters from semiflexible to  flexible limit where it acquires coil-like structure. The later state is more favourable in the presence of large fluctuations due to higher entropy.        
  
\begin{figure}
	\includegraphics[width=\columnwidth]{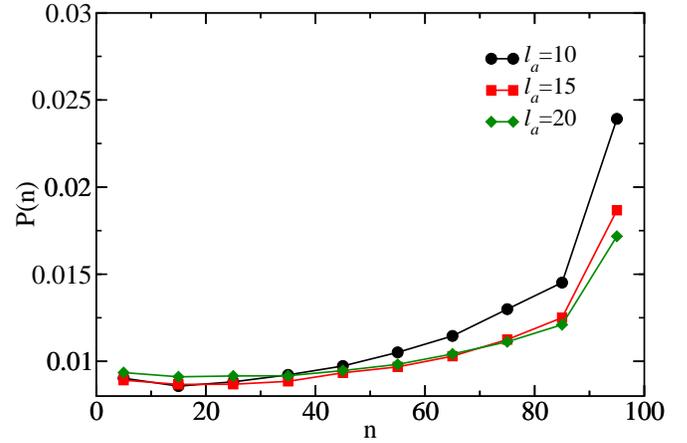}
	\caption{The distribution of active sites on the filament for the rod-like structure. Here $n=100$ corresponds to the head (front monomer) of the filament and $n=1$ is the tail.}
	\label{fig:dist_act}
\end{figure}
 
The sequential arrangement of active monomers shows  the large-scale conformational transition even at fewer number of  active monomers and small force limit. In this case, filament acquires  extended, circular, helical, and folded structures.  A helical or circular phase appears for larger separation of active monomers  
 $l_a \geq 5$ and under strong compressive force $Pe>1$. For the larger separations smaller activity is sufficient for the formation of  helical structure. This state occurs due to the competition among  drag, active  and bending forces.  A larger load applies more drag on the active monomers that requires larger force to push the filament. If  active force is weaker to drag the passive monomers then active energy is converted into  buckling of the filament.  The transition point is recognised from  the tangent-tangent correlation, which shows  crossover  from  exponential decay to oscillatory behavior \cite{giomi2010statistical}.   Large torsional order parameter also confirms twisting  and long-range ordering of  bonds  on the filament\cite{Raghu:2012}.

The load in a patterned case is very crucial  at intermediate separations.   Several interesting structures appear for large separations,i.e., $l_a \geq 5$.  Here, polymer buckles under the load in front of active monomer  that translates to all over polymer due to large separation among the active monomers. This leads to the rotation of the filament. Thus the presence of load and regular arrangement of the activity causes correlated buckling which eventually translated in circular, helical or folded structures with $Pe$. On the other hand, if the first monomer is also considered as an active  for $l_a \geq 5$ (without load) then it pulls the filament from the front end, which suppresses  buckling  between first two active monomers and causes translation of the filament.

Furthermore, to display  importance of the load on the active filaments, we perform simulations with randomly placed  active monomers termed here inhomogeneous self-propulsion. In this case, few monomers are randomly chosen to be active whose locations are also shuffled at new uncorrelated  locations with time. The shuffling is done in the fixed time interval chosen in such a way that it is larger than the longest relaxation time of the filament.   A filament acquires  structures similar to the periodic arrangement, these structures display a strong correlation with the location of the active monomers and their numbers.  

We choose to analyse here only rod-like state under inhomogeneous self-propulsion. If $R_{e}\geq 0.9Nl_0$, then it is  assumed to be  in the rod-like state. We compute the  distribution of active monomers over the filament for $R_{e}\geq 0.9Nl_0$. In other words, we identify the probability distribution of active monomers for a rod-like structure under inhomogeneous self-propulsion. 
 The distribution of active sites (Fig.~\ref{fig:dist_act})  reflects the probability of front few monomers to be active is nearly  $3$ times higher relative to others. 
This can be interpreted as presence of large number of active sites in the front drags the filament easily through 
 medium therefore it may always stay in the extended state. The structure is  similar to
 pulling a filament by a constant force. Therefore, the presence of a load is very crucial for the transition from the extended to helical, circular and folded states.

 A filament under inhomogeneous viscous drag  
 often buckles in a helical or  U-shape structures
 \cite{Raghu:2012,steinhauser2012mobility}.  Such structures  are very common in the biological systems, for example beating motion of the sperms,{\it C.elegans}, and  helical structure of  flagellum in the  bacteria\cite{Elgeti:2015,Marchetti:2013,Bechinger:2016}. In summary, we have identified that a freely swimming filament acquires helical phase in the presence of load at the front end which is so far have not been investigated in previous study. The curvature radius of the filament 
 shows  a power law behavior as, $R_\beta\sim Pe^{-\beta}$ , $\beta\sim 0.55$. The strong  buckling of the filament under the periodic arrangement of active force may provide insight in the understanding of mechanical response of the actin filaments in the presence of  molecular motors \cite{sanchez2012spontaneous,sanchez2011cilia}. 
  It would be interesting to explore the conformations of the   
  filament under the inhomogeneous self-propulsion in detail. This case may be able to provide a better comparison from the experimental systems where the medium is more complex and heterogeneous.
  
\section{Acknowledgment}
We thank funding agency DST SERB project grant no. YSS/02015/000230 for the financial support. Authors acknowledge the high performance computing facility at IISER Bhopal for the computation time.



\end{document}